\documentclass[letter, twocolumn]{jpsj2} 
\title{
Mixed-State Thermodynamics of Superconductors with Moderately Large 
Paramagnetic Effects
}

\author{Hiroto \textsc{ADACHI}
\thanks{E-mail address: adachi@mp.okayama-u.ac.jp}, 
Masanori \textsc{ICHIOKA} 
and Kazushige \textsc{MACHIDA} }

\inst{Department of Physics, Okayama University, Okayama 700-8530}

\abst{
Effects of Pauli paramagnetism on thermodynamic quantities in a vortex state, 
such as the specific heat $C$ and magnetization $M$, are studied 
using the quasiclassical Eilenberger formalism. 
We demonstrate 
that with an increase of paramagnetic depairing effect, 
the sigh of the curvature of the field dependence of $C$ changes 
from negative to positive, and that the Maki parameter 
$\kappa_2$ becomes an increasing function of temperature. 
Our results provide a natural explanation for 
the unusual field dependence of $C$ seen in CeCoIn$_5$ 
in terms of the paramagnetic effect. 
}

\kword{mixed state, thermodynamic quantity, paramagnetic effect, CeCoIn$_5$}

\begin{document}
\maketitle

 There has been much attention focused on heavy fermion superconductors, 
in particular on the newly found CeCoIn$_5$~\cite{Petrovic}. 
This material with a tetragonal symmetry exhibits a first-order 
transition~\cite{Izawa, Bianchi, Tayama, Adachi1} 
at $H_{c2}$ both for ${\bf H} \parallel c$ and ${\bf H} \perp c$. 
In a high field region for ${\bf H} \perp c$, the possible existence of 
the Fulde-Ferrell-Larkin-Ovchinnikov (FFLO) state
~\cite{Radovan, Bianchi2, Watanabe} is highly debated currently, 
because several experiments strongly indicate that 
the Pauli paramagnetic effect causes the first-order transition, 
which may lead to the FFLO state. 
The pairing symmetry of this material on either $d_{xy}$~\cite{Aoki, RI} 
or $d_{x^2-y^2}$~\cite{Izawa} is also controversial. 
One of the puzzling features concerning the pairing is that 
the specific heat $C$~\cite{Ikeda, Deguchi2, Aoki} 
obeys neither the so-called Volovik effect~\cite{Volovik} 
$C \propto \sqrt{B}$ expected for a line node case 
nor $C \propto B$ for a gapful case~\cite{Caroli, Ichioka, Nakai}, 
where $B$ is the spatially averaged internal field. 
Instead CeCoIn$_5$ shows a $C \propto B^2$-like behavior. 
We point out that a similar behavior in $C(B)$ is observed also in 
UBe$_{13}$~\cite{Ramirez}. 
It is naively expected that the exponent $\eta$ in $C \propto B^{\eta}$ 
must be $\eta \le 1$, because each vortex carries a certain 
zero-energy density of states and $C/C_{\rm n} \ge B/H_{c2}$, 
where $C_{\rm n}$ is the normal state specific heat.
Thus, the observed $C \propto B^2$-like behavior is highly unusual and 
a puzzling issue to be solved. 

In a clean singlet type-II superconductor 
under a magnetic field, 
there are two microscopic energy scales~\cite{Fetter, Vekhter, Carbotte} 
for the depairing effects: 
the Zeeman energy $\mu B$ for the paramagnetic depairing 
and the orbital depairing energy ${\cal E}_B= v_F/r_B$ 
associated with the Doppler shift, where 
$v_F$ is a Fermi velocity, $r_B= \sqrt{\phi_0/2 \pi B}$ 
with $\phi_0$ being the flux quantum. 
In materials with a small Fermi velocity such as heavy fermion superconductors, 
the role of the Zeeman energy compared to ${\cal E}_B$ cannot be neglected when 
discussing the thermodynamic properties. 
As for the phenomena near $H_{c2}$, the effects of Pauli paramagnetism 
on the mixed state have been studied by Adachi and Ikeda 
using the nonlocal Ginzburg-Landau (GL) theory~\cite{Adachi1}. 
To describe the thermodynamic quantities in lower fields 
we use the quasiclassical Eilenberger formalism~\cite{Eilenberger}. 

In this Letter, we clarify how 
Pauli paramagnetism affects the mixed-state thermodynamics 
using the quasiclassical Eilenberger formalism. 
Using the approximate method given in ref.~\citen{Adachi2}, 
we demonstrate that in a superconductor with moderately 
large paramagnetic effects under a magnetic field, 
the specific heat varies roughly quadratically with the field, $C \propto B^2$, 
consistent with the experimental data in CeCoIn$_5$~\cite{Ikeda, Deguchi2, Aoki}. 
At a lower temperature with a stronger paramagnetic effect, 
our result shows a magnetization jump at $H_{c2}$, which is also observed 
in CeCoIn$_5$~\cite{Tayama}. 
We give a simple thermodynamic argument to relate the positive curvature 
of $C(B)$ to the magnetization behavior under the paramagnetic effect, 
giving rise to the internal consistency between these quantities. 
Near $H_{c2}$, our numerical solution coincides, of course, with the results of 
the nonlocal GL theory given in ref.~\citen{Adachi1}. 

First, let us briefly introduce our approach. 
We start with the Eilenberger equation~\cite{Eilenberger} 
incorporating the paramagnetic effect~\cite{Klein, Kita} 
($\hbar= k_{B}=1$); 
\begin{equation}
\Big( 2 \varepsilon_\mu + {\rm i} {\bf v \cdot \Pi} \Big) 
f(\varepsilon_{\mu}, {\bf p}, {\bf r} )
=
2 g(\varepsilon_{\mu}, {\bf p}, {\bf r} ) w_{\bf p}\Delta({\bf r}), 
\label{eq:Eilen}
\end{equation}
where $f(\varepsilon_{\mu}, {\bf p}, {\bf r} )$, 
$f^\dag(\varepsilon_{\mu}, {\bf p}, {\bf r} )= 
[f(\varepsilon_{\mu}^*, -{\bf p}, {\bf r} )]^*$, 
and $g(\varepsilon_{\mu}, {\bf p}, {\bf r} )= 
\sqrt{1-f f^\dag}$ 
are the quasiclassical Green's functions. 
Here, ${\bf v}=v_F {\bf \hat{p}}$ is a Fermi velocity, 
${\bf \Pi}=-{\rm i} {\bf \nabla}+ (2 \pi/ \Phi_0){\bf A}$ is a 
gauge invariant gradient, 
$T_c$ is a transition temperature at a zero field. 
The fermionic Matsubara frequency $\varepsilon_n= 2 \pi T(n+1/2)$ 
combined with the (renormalized) Zeeman energy is denoted as 
$\varepsilon_\mu= \varepsilon_n- {\rm i} \mu B$, 
and the gap function is expressed as 
$\Delta_{\bf p}({\bf r})= w_{\bf p} \Delta({\bf r})$ 
with the pairing function $w_{\bf p}$ and the pair field $\Delta({\bf r})$. 
In the following, the pairing state of 
a $d$-wave ($w_{\bf p}=2\sqrt{2} \hat{p}_x \hat{p}_y $) 
or $s$-wave ($w_{\bf p}=1$) is assumed. 
Since the detailed shape of the Fermi surface 
does not change our main results essentially, 
we consider a quasi-two-dimensional isotropic Fermi surface. 
In the mixed state containing field-induced vortices, 
the pair field $\Delta$ can be expanded into each Landau level: 
\begin{equation}
  \Delta = \Delta_0 \sum_{N =0}^{N_{\rm max}} d_N \psi_N, 
\end{equation}
where 
$\psi_N = \sum_{m=-\infty}^{\infty} C_m 
\frac{ H_N( y+ \nu m)}{\sqrt{2^N N!}} {\rm e}^{-(y+ \nu m)^2/2-{\rm i}\nu m x }$,
$\Delta_0=1.764 T_c$, 
$C_m= (\nu^2 / \pi)^{1/4}  {\rm e}^{- {\rm i} \pi \zeta m^2}$, 
$H_N$ the $N$-th Hermite polynomial, 
and the lengths are measured in units of $r_B$. 
The real constants $\zeta$ and $\nu$ are set to be 
$\zeta=1/2$ and $\nu= {(3 \pi^2)}^{1/4}$, assuming 
a triangular vortex lattice.  
The difference in vortex configuration gives minor contributions. 
To solve eq. (\ref{eq:Eilen}), we adopt the approximate method given 
in ref.~\citen{Adachi2}, which gives  
\begin{equation}
 f = 2 g w_{\bf p} \int_0^{\infty} d \rho \; 
{\rm e}^{-2 \varepsilon_\mu \rho} 
\Big( \Delta_0 \sum_{N =0}^{N_{\rm max}} \alpha_N d_N \Big).   
\end{equation}
Here, the quantity $\alpha_N$ is given by 
$ \alpha_N = 
 \sum_{m} C_m
 \frac{ H_N( y + \nu m- {\rm Re}\, \lambda)}{\sqrt{2^N N!}} 
{\rm e}^{-(|\lambda|^2- \lambda^2)/4}{\rm e}^{-(y+\nu m- \lambda)^2/2-{\rm i}\nu m x }$, 
and $\lambda=(v_y+ {\rm i}v_x) \rho/r_B$. 
The above procedure uses an approximation 
similar to that used by Pesch~\cite{Pesch}, in which 
the operator ${\bf \Pi}$ does not pick up the spatial variation in 
$g$, but only that in $\Delta$. 
For extreme type-II materials with a large GL parameter $\kappa$
(in CeCoIn$_5$ $\kappa \gg 10$), 
this approximation provides a fairly good description 
for almost all experimentally relevant fields. 
Note that in contrast to the method by Pesch, our method 
can reproduce the first non-Gaussian term of the nonlocal 
GL free energy~\cite{Adachi1}. 
Recently, a simplified version of our method has been 
widely used~\cite{BPT-approx} in slightly different contexts, 
in which the spatial degree of freedom is integrated out in advance. 

The expansion coefficients $\{ d_N \}$ are determined by the following 
gap equation 
\begin{equation}
  \Big[ \ln (\frac{T}{T_{c}}) 
+  \sum_{n \ge 0} \frac{2 \pi T}{ \varepsilon_n}  \Big] d_N 
=
 \frac{\pi T}{\Delta_0} \sum_{n \ge 0}
  \overline{ \psi_N^* \langle  w_{\bf p}^* \big( f+ (f^{\dag})^* \big) 
\rangle},  
\end{equation}
where $\langle \cdots \rangle$ and $\overline{\cdots}$ indicate the 
Fermi surface and spatial average, respectively. 
The strength of the paramagnetic effect in this Letter is 
measured using $\alpha_{\rm para} \equiv \mu H_{\rm orb}/2 \pi T_c$, where 
$H_{\rm orb}=0.561 \Phi_0/2 \pi \xi_0^2$ is the two-dimensional orbital limiting field 
and $\xi_0= v_F/2 \pi T_c$. 
If we define another parameter introduced by Maki as 
$\alpha_{\rm M}= \sqrt{2} H_{\rm orb}/H_P \; (H_P=\Delta_0/\sqrt{2})$
~\cite{Fetter, Maki}, 
these two parameters are related as $\alpha_{\rm M}= 7.12 \alpha_{\rm para}$. 
Throughout this Letter, we limit our discussion to the parameter region 
$\alpha_{\rm para} \le 0.6$ so that the possibility of 
either the vortex state 
formed by odd Landau levels~\cite{RI, Houzet, Klein} or 
the FFLO vortex state~\cite{RI} is safely excluded, 
and we consider only the lowest Landau level. 
Furthermore, we define $\kappa$ as $\kappa^{-2}=4 \pi N(0) \Delta_0^2/H_{\rm orb}^2$, 
where $N(0)$ is the density of states at the Fermi surface. 
\begin{figure}[t] 
\scalebox{0.5}[0.5]{\includegraphics{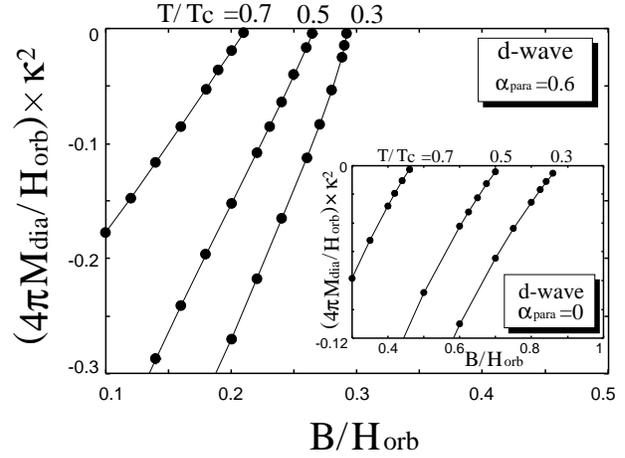}}
\caption{
Field dependence of $M_{\rm dia}$ in $d$-wave superconductor 
with $\alpha_{\rm para}=0.6$. 
Inset: corresponding data for $\alpha_{\rm para}=0$. 
}
\label{fig1}
\end{figure}

To confirm whether our method can correctly capture the 
paramagnetic effect in the mixed state, 
we first study the magnetization. 
In the presence of the paramagnetic effect, 
the mixed-state magnetization is decomposed into 
the diamagnetic part $M_{\rm dia}$ and the paramagnetic part 
$M_{\rm para}$~\cite{Kita, Klein}:
\begin{eqnarray}
 4 \pi M_{\rm dia}
&=&
 -\frac{2 \pi^2 N(0)}{B} 
T \sum_{n \ge 0} 
\Big \langle 
\overline{
\frac{g(f^{\dagger}(w_{\bf p}\Delta)+ f (w_{\bf p}\Delta)^*) }{1+g}
} \nonumber \\
&&  \hspace{2cm} - 2 \varepsilon_n \overline{(1-g)}   \Big\rangle + {\rm c.c.}, \\
 4 \pi M_{\rm para}
&=&
 4 \pi M_{\rm n} \Big( 
 1+ \big( \frac{2 \pi T}{\mu B} \big) {\rm Im} 
\sum_{n \ge 0}  \overline{ \langle g \rangle }
\Big), 
\end{eqnarray} 
where $M_{\rm n}=2 \pi \mu^2 N(0) B$ is the Pauli paramagnetism 
in the normal state.
In the clean limit without paramagnetism, 
the Maki parameter 
$\kappa_2= \sqrt{ [2 \beta_A \; d(4 \pi M_{\rm dia})/dH]^{-1}+ 0.5}|_{Hc2}$ 
with $\beta_A= 1.1596$ 
is known to be a decreasing function of the temperature $T$~\cite{Eilenberger2, Kita2}. 
In the high-$\kappa$ case, 
we approximately have 
$\kappa_2 \propto [\frac{d}{dB}(4 \pi M_{\rm dia}) ]^{-1/2}_{B=H_{c2}}  $, 
and the magnetization slope at $H_{c2}$ gives a rough estimate of $\kappa_2$. 
The inset of Fig.~\ref{fig1} shows the field dependences of $M_{\rm dia}$ 
in a $d$-wave superconductor without the paramagnetic effect 
($\alpha_{\rm para}=0$). 
We can see that $\kappa_2$ is a slowly decreasing function of 
$T$ in agreement with the results shown in refs.~\citen{Eilenberger2} 
and~\citen{Kita2}. 
Corresponding data with the paramagnetic effect 
($\alpha_{\rm para}=0.6$) are shown in the main panel of Fig.~\ref{fig1}. 
Contrary to the purely diamagnetic case, $\kappa_2$ in this case 
is an increasing function of $T$ consistent with the observation 
in CeCoIn$_5$~\cite{Ikeda, Deguchi2}. 
Our result should be compared with that of the dirty-limit analysis~\cite{Maki}, 
where $\kappa_2$ was shown to be an increasing function of $T$ 
due to the paramagnetic effect. 
 
\begin{figure}[b]
\scalebox{0.5}[0.5]{\includegraphics{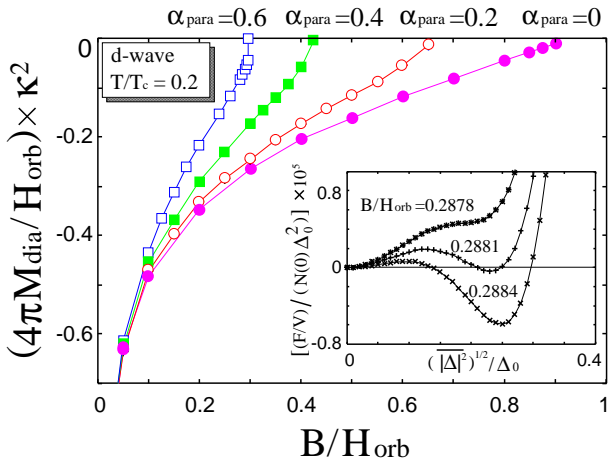}}
\caption{
Field dependence of $M_{\rm dia}$ at $T/T_c = 0.2$ 
in $d$-wave superconductor 
for $\alpha_{\rm para}=0$, $0.2$, $0.4$, and $0.6$. 
Inset: free energy near $H_{c2}$ vs pair field strength 
for $\alpha_{\rm para}=0.6$. 
The expression used for 
the free energy is the same as eq. (4) in ref.~\citen{Kita}. 
}
\label{fig2}
\end{figure}

Next, we study the magnetization at a lower temperature 
where a first-order transition at $H_{c2}$ occurs~\cite{Houzet, Adachi1}. 
Figure~\ref{fig2} shows the field dependences of $M_{\rm dia}$ 
at $T/T_{c}=0.2$ in a $d$-wave superconductor. 
In a low field region, the paramagnetic effect is irrelevant and all the data 
for a different $\alpha_{\rm para}$ value merge into a single curve. 
On the other hand, the slope $d M_{\rm dia}/d B$ increases 
with increasing $\alpha_{\rm para}$ near $H_{c2}$, 
and for $\alpha_{\rm para}=0.6$, we find a 
discontinuous jump in magnetization at $H_{c2}$ 
reflecting the first-order transition within the mean-field approximation. 
Indeed, if we plot the free energy against the pair field strength 
near $H_{c2}$ (inset of Fig.~\ref{fig2}), 
the free energy has a minimum at a nonzero value of 
$(\overline{|\Delta|^2})^{1/2}$, 
implying a discontinuous jump in corresponding thermodynamic quantities. 
This explains the nearly discontinuous transition at $H_{c2}$ 
observed in CeCoIn$_5$~\cite{Izawa, Bianchi, Tayama}. 
We note here that we have also obtained 
essentially the same results also for the $s$-wave case.

\begin{figure}[b,h]
\scalebox{0.45}[0.45]{\includegraphics{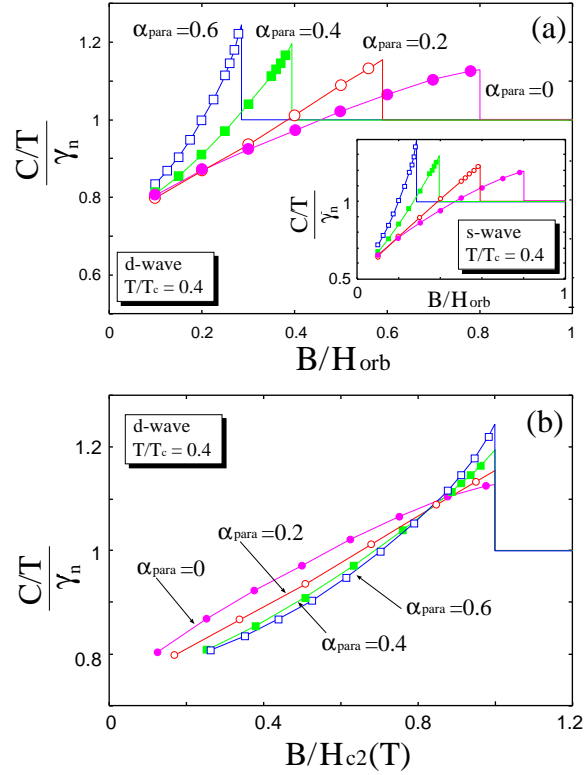}}
\caption{
(a) Field dependence of specific heat at $T/T_{c}=0.4$  
for $d$-wave case. Inset: corresponding data for $s$-wave case. 
Here, $\gamma_{\rm n}=2 \pi^2 N(0)/3$. 
(b) Data in (a) are plotted as a function of the normalized field $B/H_{c2}(T)$. 
}
\label{fig3}
\end{figure}

Now, we consider the specific heat $C$. 
To obtain $C$ numerically, we first calculate the entropy difference between 
the normal and mixed states using~\cite{Kita} 
\begin{eqnarray}
  \frac{S}{V}- \frac{S_{\rm n}}{V} &=& 
\frac{N(0)}{T} \Bigg( - \overline{|\Delta|^2} 
+{\rm Re} \; 2 \pi T \sum_{n \ge 0}  
\Big\langle 
 2 \varepsilon_n \overline{(g-1)} \nonumber \\ 
&&\hspace{1.2cm}
+\overline{ \frac{f^{\dagger}(w_{\bf p}\Delta)+ f (w_{\bf p}\Delta)^* }{1+g} }
   \Big\rangle   \Bigg),  
\end{eqnarray}
where $S_{\rm n}/V= 2 \pi^2 N(0)T/3$ is the entropy in the normal state. 
Then we perform a polynomial interpolation through the numerical data, 
and obtain $C$ using $C= T dS/dT$. 
Figure~\ref{fig3}(a) shows the field dependence of $C/T$ at $T/T_{c}=0.4$ 
in a $d$-wave superconductor. 
In the absence of the paramagnetic effect ($\alpha_{\rm para}$=0), 
the field dependence of $C/T$ has a negative curvature, 
which coincides with the results shown in refs.~\citen{Ichioka} and~\citen{Nakai}. 
With increasing $\alpha_{\rm para}$, however, 
the sign of the curvature changes from negative to positive. 
This is more evident in Fig.~\ref{fig3}(b), 
where the same data are plotted against the normalized field $B/H_{c2}(T)$. 
This field dependence is consistent with the observation in 
CeCoIn$_5$~\cite{Ikeda, Deguchi2, Aoki}. 
It is worth noting that this behavior can be seen irrespective 
of the structure of the pairing function. 
Actually, the same behavior is seen for the $s$-wave case~\cite{Comm}, 
as shown in the inset of Fig.~\ref{fig3}(a). 
Also we would like to mention that although our result is based on an 
approximate solution and should be understood 
as a semi-quantitative one at most, 
an extensive numerical calculation~\cite{IMcomm} 
using the so-called 
explosion method supports the conclusion. 

\begin{figure}[b,h]
\scalebox{0.5}[0.5]{\includegraphics{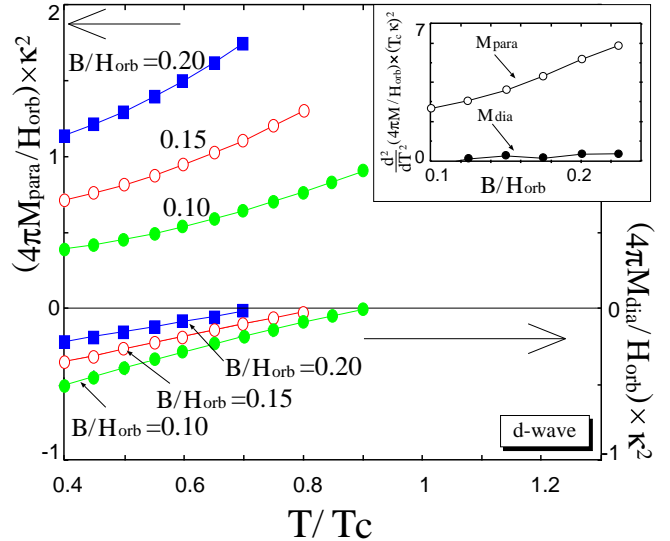}}
\caption{
Temperature dependences of $M_{\rm para}$ and $M_{\rm dia}$ for $\alpha_{\rm para}=0.6$ 
at several fields.  
Inset: field dependences of $\partial^2 M_{\rm para}/ \partial T^2 $ 
and $\partial^2 M_{\rm dia}/ \partial T^2 $ at $T/T_{c}=0.4$ 
for $\alpha_{\rm para}=0.6$. 
} 
\label{fig4}
\end{figure}

To understand the microscopic origin of the behavior $C(B) \propto B^2$, 
it is convenient to relate the magnetization with $C(B)$ through 
a thermodynamic Maxwell's relation, 
\begin{equation}
  \frac{ \partial }{\partial B}\; \left( \frac{C(B)}{T} \right)
=
\frac{ \partial^2 }{ \partial T^2} \; \Big( M(T) \Big), 
\end{equation}  
indicating that the $T^2$-coefficient of the magnetization 
gives the slope of $C(B)/T$. 
Since the magnetization can be divided into the paramagnetic ($M_{\rm para}$) 
and diamagnetic ($M_{\rm dia}$) parts, 
we can separately discuss the corresponding contribution to $C(B)$ 
through the above relationship. 
Figure~\ref{fig4} shows the temperature dependences of $M_{\rm dia}$ and $M_{\rm para}$ 
for $\alpha_{\rm para}= 0.6$ at several fields. 
It is easily seen that $M_{\rm para}$ changes like $T^2$ and this coefficient 
increases with increasing magnetic field, 
whereas $M_{\rm dia}$ seems to be almost $T$-linear and has a small 
$T^2$-coefficient. 
This is clearer in the inset of Fig.~\ref{fig4}, where 
the field dependences of $\partial^2 M_{\rm para} / \partial T^2$ 
and $\partial^2 M_{\rm dia}/  \partial T^2$ are shown; 
$ \partial^2 M_{\rm para}/ \partial T^2 $ is almost $B$-linear and 
gives the $B^2$-dependence of $C(B)$, 
while $\partial^2 M_{\rm dia}/ \partial T^2 $ is almost 
field-independent and does not give a large field dependence. 
Thus, the main source of the behavior $C(B) \propto B^2$ 
is not the vortex contribution but the paramagnetic depairing of the Cooper pair 
that lowers $H_{c2}$ considerably. 
It is interesting to point out that $C(B)/T \propto \sqrt{B}$ 
for a nodal gap case ($\propto B$ for a gapful case) at low temperatures 
indicates $\beta(B) \propto 1/\sqrt{B}$ 
($\propto$ const), where $M(T)= M(0)+ \alpha(B) T+ \beta (B) T^2+ O(T^3)$. 
Although there is no accurate measurement for $M(T)$ to check it, 
the existing data of YNi$_2$B$_2$C~\cite{Song} and MgB$_2$~\cite{Lee} 
seems to support it. 
It is desirable to confirm this internal consistency for other superconductors.

We briefly discuss the data of Sr$_2$RuO$_4$, 
which is believed to be a prime candidate for the triplet pairing. 
For ${\bf H} \perp c$, $C(B)/T$ shows a strong positive curvature 
and changes smoothly into a $\sqrt{B}$ behavior as {\bf H} rotates towards the 
$c$-direction~\cite{Deguchi1}, 
reminiscent of the change with $\alpha_{\rm para}$ shown in Fig.~\ref{fig3}. 
This may be understandable if we assume either that 
the $d$-vector in Sr$_2$RuO$_4$ is strongly locked within the basal plane~\cite{Sauls} 
or singlet pairing symmetry.  
Upon rotating the field direction, the relative weight $H_{\rm orb}$ 
to the paramagnetic effect $\mu B$, or $\alpha_{\rm para}$, changes, 
and $C(B)/T$ should show positive to negative curvatures. 
Note that~\cite{Tenya} 
(i) the in-plane magnetization curve exhibits a jump near $H_{c2}$ at low 
temperatures, 
(ii) the Maki parameter $\kappa_2$ is an increasing function of temperature, 
and (iii) $H_{c2}$($\perp c$) is strongly suppressed; 
these behaviors are similar to those of CeCoIn$_5$. 

In conclusion, we have studied the mixed-state thermodynamics of a 
superconductor with moderately large paramagnetic effects. 
Our numerical calculation based on the quasiclassical Eilenberger formalism 
revealed that with increasing paramagnetic effect, 
(i) the Maki parameter $\kappa_2(T)$ becomes indeed an increasing function of $T$ 
and (ii) a $C(B) \propto B^2$ behavior appears. 
It can be said that the field dependence of the specific heat and magnetization 
is a good probe for measuring the strength of 
paramagnetic effects in type-II materials. 
Moreover, our demonstration provides a natural explanation for the 
$C(B) \propto B^2$ behavior observed in CeCoIn$_5$ 
in terms of the paramagnetic effect. 

After this work was completed, we were informed~\cite{RIcomm} 
that a $C(B) \propto B^2$ behavior is a universal phenomenon 
for a system on the verge of exhibiting a first-order transition of 
superconductivity; 
thus, it can also arise from a strong coupling effect 
such as antiferromagnetic fluctuations. 

We acknowledge fruitful discussions with T. Sakakibara, R. Ikeda, K. Tenya, 
T. Tayama, and K. Deguchi.

\bibliography{basename of .bib file}

\begin{thebibliography}{99}
\bibitem{Comm}
Our calculation was performed at relatively high temperatures, 
covering the behavior that depends on the pairing symmetry.  
In a zero-temperature limit, we observe a 
$C \propto \sqrt{B}$-like behavior~\cite{IMcomm} at a low field region 
for a line node case.
\end{thebibliography}

\end{document}